\pdfoutput=1
\documentclass[journal,twoside,web]{ieeecolor}
\usepackage{generic}
\usepackage{cite}
\usepackage{amsmath,amssymb,amsfonts}
\usepackage{algorithmic}
\usepackage{graphicx}
\usepackage{algorithm,algorithmic}
\usepackage{hyperref}
\hypersetup{hidelinks=true}
\usepackage{textcomp}
\usepackage{url}
\usepackage{xcolor}
\usepackage{float}
\usepackage{hyperref}
\usepackage{stfloats}
\usepackage{amssymb}
\usepackage{latexsym}
\usepackage{amsmath}
\usepackage{makecell}
\usepackage{booktabs}
\usepackage{arydshln}
\usepackage{array}
\usepackage{multirow}
\usepackage{upgreek}
\usepackage{pifont}
\usepackage{booktabs}
\usepackage{multirow}
\usepackage{pifont} 
\usepackage{graphicx}
\usepackage{array}
\usepackage{bbding}

\def\BibTeX{{\rm B\kern-.05em{\sc i\kern-.025em b}\kern-.08em
    T\kern-.1667em\lower.7ex\hbox{E}\kern-.125emX}}
\begin{document}
\title{Medical Image Fusion for High-Level Analysis: A Mutual Enhancement Framework for Unaligned PAT and MRI}
\author{Yutian Zhong, Jinchuan He, Zhichao Liang, Shuangyang Zhang, Qianjin Feng, Lijun Lu, and Li Qi 
\thanks{This work was supported in part by National Natural Science Foundation of China (62371220), Guangdong Basic and Applied Basic Research Foundation (2021A1515012542, 2022A1515011748), and Guangdong Pearl River Talented Young Scholar Program (2017GC010282). (\textit{Corresponding authors: Lijun Lu; Li Qi.})}
\thanks{The authors are with School of Biomedical Engineering, Southern Medical University, Guangzhou, Guangdong, China; Guangdong Provincial Key Laboratory of Medical Image Processing, Southern Medical University, Guangzhou, Guangdong, China; and Guangdong Province Engineering Laboratory for Medical Imaging and Diagnostic Technology, Southern Medical University, Guangzhou, Guangdong, China. (e-mail: ytzhong.smu@qq.com; 2512549169@qq.com; 2153136120@qq.com; syzhang@smu.edu.cn; fengqj99@smu.edu.cn; lulijun@smu.edu.cn; qili@smu.edu.cn).}}

\maketitle

\begin{abstract}
Photoacoustic tomography (PAT) offers optical contrast, whereas magnetic resonance imaging (MRI) excels in imaging soft tissue and organ anatomy. The fusion of PAT with MRI holds promising application prospects due to their complementary advantages. Existing image fusion have made considerable progress in pre-registered images, yet spatial deformations are difficult to avoid in medical imaging scenarios. More importantly, current algorithms focus on visual quality and statistical metrics, thus overlooking the requirements of high-level tasks. To address these challenges, we propose an unsupervised fusion model, termed PAMRFuse+, which integrates image generation and registration. Specifically, a cross-modal style transfer network is introduced to simplify cross-modal registration to single-modal registration. Subsequently, a multi-level registration network is employed to predict displacement vector fields. Furthermore, a dual-branch feature decomposition fusion network is proposed to address the challenges of cross-modal feature modeling and decomposition by integrating modality-specific and modality-shared features. PAMRFuse+ achieves satisfactory results in registering and fusing unaligned PAT-MRI datasets. Moreover, for the first time, we evaluate the performance of medical image fusion with multi-organ instance segmentation. Extensive experimental demonstrations reveal the advantages of PAMRFuse+ in improving the performance of medical image analysis tasks.
\end{abstract}


\section{Introduction}
\label{sec:introduction}

In biomedical imaging, photoacoustic tomography (PAT) and magnetic resonance imaging (MRI) represent two advanced imaging modalities. PAT combines the high contrast of optical imaging with the deep imaging range of ultrasound, enabling the visualization of fine structures and tissue characteristics \cite{b1}--\cite{b2}. In contrast, MRI utilizes magnetic fields and radio waves to provide excellent soft tissue contrast \cite{b5}. Although PAT can provide functional and molecular information about tissues, it is limited by tissue optical scattering and lacks soft tissue contrast. Conversely, MRI can offer good soft tissue information, but its temporal resolution is limited. This complementarity motivates us to integrate the unique information between PAT and MRI images to obtain more comprehensive information.

\begin{figure*}[!h]
\centering
\includegraphics[scale=0.6]{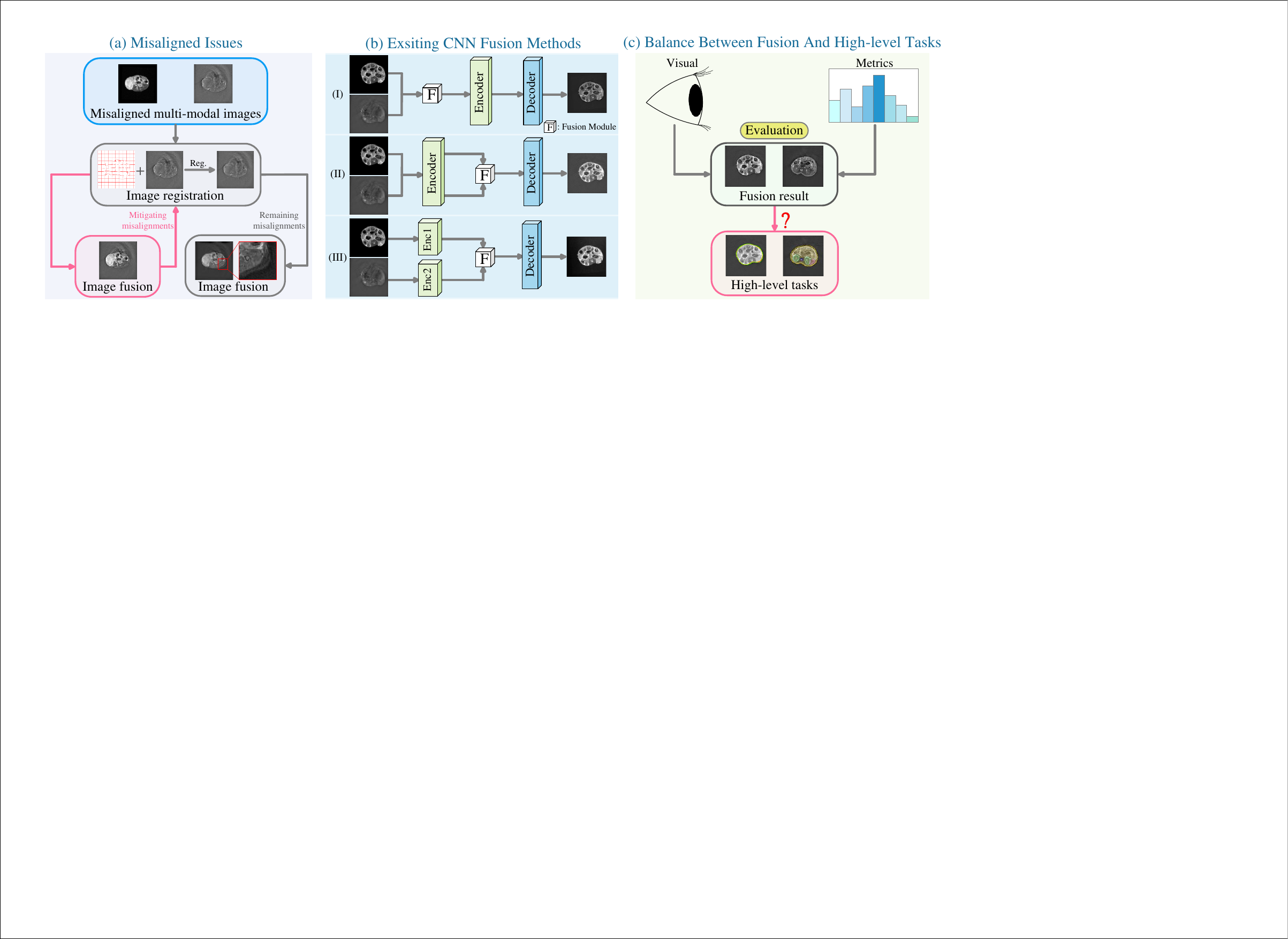}
\caption{Image fusion methods face some pressing challenges. (a) Misalignment of source images leads to artifacts in the fusion results. Reg.=Registration. (b) Existing fusion methods fail to produce higher-quality fusion images. (c) In terms of evaluation, current fusion methods overlook adaptability to advanced visual tasks.}
\end{figure*}

Many existing approaches for integrating PAT and MRI information primarily rely on hardware or registration techniques \cite{b7}--\cite{b8}. However, these methods have yet to achieve true information fusion. The goal of image fusion is to extract valuable information from each image and combine them into a single image with more comprehensive information and improved visualization. Due to its practical applicability, image fusion finds wide-ranging applications in the medical imaging domain \cite{b9}, including CT-MRI, CT-PET, MRI-PET, etc. Our group recently introduced the PAMRFuse network \cite{b10} which achieves PAT and MRI fusion. Despite the capability of the PAMRFuse network to generate good fusion images, there remain pressing challenges in the field of PAT and MRI image fusion.

Firstly, when the source images are misaligned, this leads to noticeable artifacts in the fusion results, as shown in Fig. 1 (a). Therefore, fusion methods need to employ image registration as a preprocessing step to alleviate misalignments. Unfortunately, the diversity between different modalities poses significant challenges to improving registration accuracy. In this case, registration and fusion are usually treated as two independent issues. Misaligned regions in the input images lead to repeated prominent structures, while accurate registration facilitates gradient sparsity. Thus, gradient sparsity may serve as a criterion for feedback to improve registration accuracy, and accurate alignment data will further enhance fusion results.

Secondly, regarding the challenges of image fusion, methods based on convolutional neural networks (CNN) for feature extraction and fusion have many shortcomings. As a basic model, CNN is difficult to control and interpret, leading to inadequate extraction of cross-channel features. For example, in Fig. 1 (b), the shared encoder in (i) and (ii) fails to distinguish modality-specific features, while the private encoder in (iii) overlooks features shared between modalities.

Thirdly, as shown in Fig. 1 (c), existing fusion algorithms tend to prioritize enhancing visual quality and evaluation metrics, while neglecting the adaptability of fused images for high-level tasks. To ensure the accuracy and reliability of these tasks, the fused images should not only have good visual quality but also retain the semantic information and structural features of the original images. Therefore, necessitates a balance between improving the visual quality of fused images and considering the adaptability and effectiveness for various high-level tasks.

High-level tasks can be considered as the concrete utilization of information contained within images, which is indispensable in medical imaging analysis. Specifically, in the application of PAT, tissue or organ segmentation is often a prerequisite for the region of interest analysis and molecular imaging, where effective segmentation assists in computing biomarker concentrations and estimating the effects of internal light attenuation, thereby achieving quantitative imaging \cite{b28}. Unfortunately, due to insufficient resolution and contrast of existing PAT imaging technologies, the segmentation performance is suboptimal \cite{b25}. There is a pressing need for an effective auxiliary technique that preserves the advantages of PAT images while meeting the requirements of advanced visual tasks.

To address the limitations of previous work and unexplored issues, we propose for the first time the use of image fusion to enhance advanced visual tasks in PAT, and achieve multimodal PAT-MRI image synthesis, registration, and fusion within a mutually reinforcing framework. We introduce an unsupervised network to achieve this, named PAMRFuse+. We introduce a cross-modal generation-registration paradigm for the alignment of images. Subsequently, PAMRFuse+ explores a rational paradigm to address challenges in feature extraction and fusion by proposing a dual-branch feature decomposition fusion (DFDF) model. We evaluate the proposed method on unaligned PAT-MRI datasets, achieving excellent performance in both image registration and fusion simultaneously. More crucially, we assess the effects of image fusion in advanced visual tasks such as multi-organ instance segmentation, where PAMRFuse+ shows its potential to facilitate performance in advanced visual tasks.

The main contributions are summarized as follows:

1. We propose for the first time the use of image fusion to enhance advanced analysis tasks in PAT and introduce a highly robust unsupervised PAT and MRI image fusion framework.

2. We develop a generation-registration paradigm and mutual reinforcement of the registration and fusion tasks to overcome the difficulty of cross-modal source image alignment.

3. For the fusion network, we propose a dual-branch transformer framework for extracting and fusing modality-specific and modality-shared features.

4. Qualitative and quantitative results in PAT and MRI images exemplify the excellent alignment and fusion performance of PAMRFuse+. Importantly, we employ an advanced visual task to assess the performance of image fusion. 

The preliminary version of this work is PAMRFuse \cite{b10}. Compared to the initial version, the specific technical improvements are as follows:

1. In PAMRFuse, the registration and fusion tasks are independent. PAMRFuse must "tolerate" rather than "counteract" the misalignments post-registration. PAMRFuse+ treats the synthesis, registration, and fusion processes as an integrated whole to mutually enhance the performance of each task.

2. PAMRFuse uses GAN as an image synthesis network. Due to the lack of one-to-one correspondence in multimodal data, its performance is limited. PAMRFuse+ adopts CycleGAN's cycle consistency method, removing the restriction of image translation.

3. Compared to the coarse and single-stage image registration network in PAMRFuse, PAMRFuse+ uses a multi-stage registration network to predict the deformation field, which further improves registration accuracy.

4. In terms of image fusion, PAMRFuse uses a shared encoder for all inputs, making it difficult to distinguish modality-specific features. PAMRFuse+ designs a separate encoder for each modality to address the challenges of feature extraction.

5. PAMRFuse only considers enhancing the visual quality and evaluation metrics of the fused images. PAMRFuse+ not only achieves good visual quality but also preserves the semantic information of the original images. Its potential to enhance the performance of advanced visual tasks is demonstrated on multi-organ instance segmentation tasks.

\section{Methods}

\subsection{Cross-modality Style Transfer}
One inherent characteristic of PAT images is the emphasis on texture details rather than sharp geometric structures. To mitigate the low contrast issue caused by scattering in PAT images, generating pseudo-MRI images with clear structural information is more advantageous for monomodal registration. Therefore, we propose a cross-modal style transfer network (P2M) to extract accurate geometric structures during the generation of pseudo-MRI images $I_{\overset{\sim}{MR}}$ from PAT images $I_{PA}$, as depicted in Fig. 2 (a). This network inherits the cycle-consistent learning approach from CycleGAN \cite{b31}. However, unlike CycleGAN, we aim to design a specific learning strategy controlled by perceptual style transfer constraints to further optimize the generation of pseudo-MRI images with sharp structures.

\subsection{Multi-level Refinement Registration}
As illustrated in Fig. 2 (b), we employ a multi-level registration network to predict the deformation field between MRI images and pseudo-MRI images, and reconstruct the registered PAT image. MLR consists of a shared multi-layer feature extraction module, two coarse-to-fine deformation field estimation modules, and a resampling layer. Each coarse-to-fine deformation field estimation module includes a coarse deformation field estimation (CDFE) module $\mathcal{M}_{\mathcal{C}}$ and a refined deformation field estimation (RDFE) module $\mathcal{M}_{\mathcal{R}}$. The coarse deformation field is predicted as follows:

\begin{equation} 
\varphi_{c}^{k} = \mathcal{M}_{\mathcal{C}}\left( {\mathcal{F}^{k}\left( {I_{\overset{\sim}{MR}},I_{MR}} \right)} \right)
\end{equation}

\noindent where $\mathcal{F}^{k}$ represents the $k$-th level feature extraction module. 

The estimation formula for the refined deformation field is:

\begin{equation} 
\varphi_{r}^{k} = \mathcal{M}_{\mathcal{R}}\left( \varphi_{c}^{k} \right) \oplus \varphi_{c}^{k}
\end{equation}

Assuming the layer feature extraction module consists of $K$ layers, when $k = K$, the final estimated deformation field is $\phi = {- \varphi}_{r}^{k}$. Finally, we use the resampling layer to reconstruct the registered PAT images:

\begin{equation} 
I_{PA}^{reg} = I_{PA}{^\circ}(\phi)
\end{equation}

\noindent The operator ${^\circ}$ denotes the spatial transformation.

\begin{figure}[ht]
\centering
\includegraphics[scale=0.45]{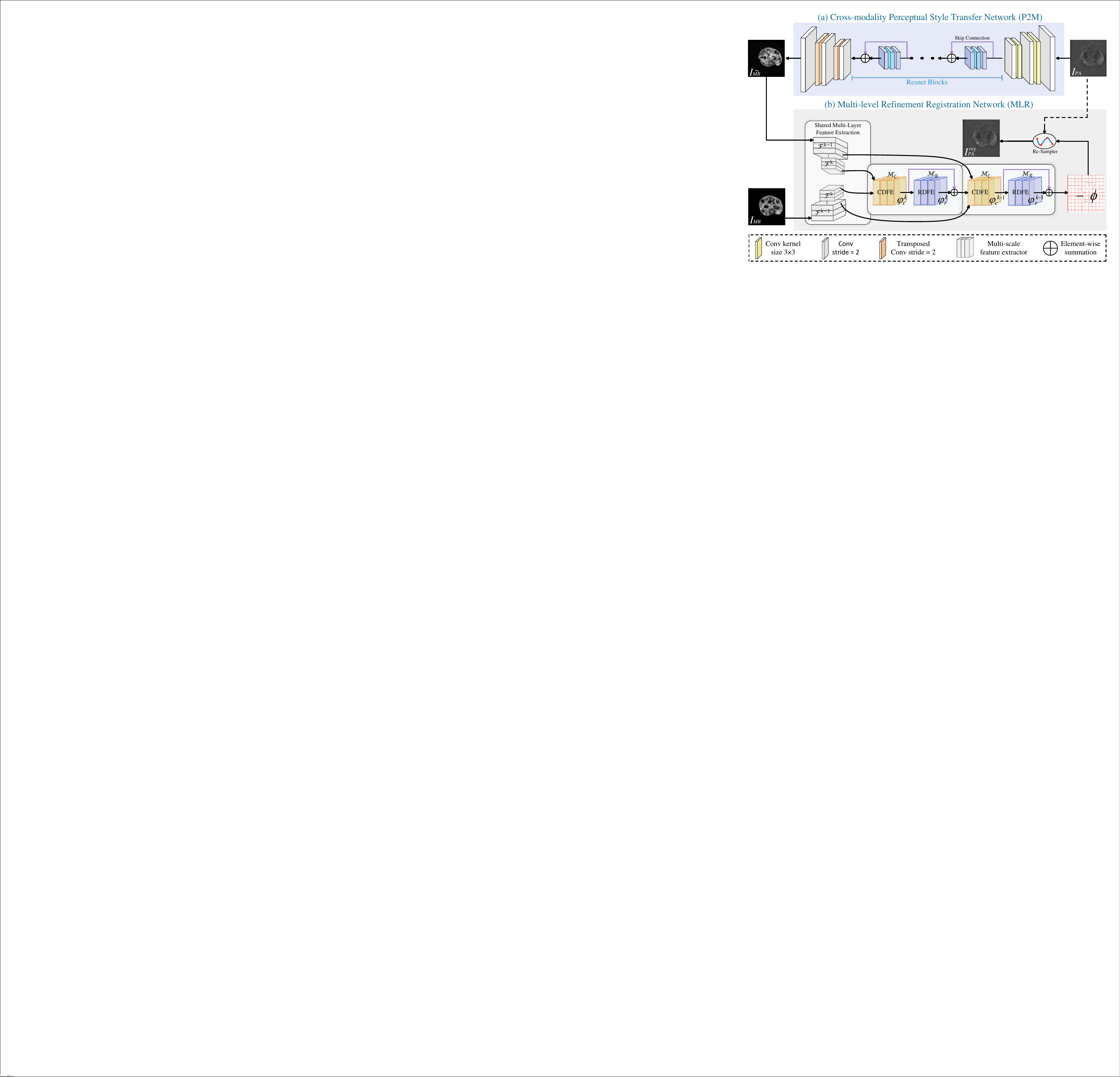}
\caption{The workflow of the cross-modal synthesis-registration paradigm.}
\end{figure}

\begin{figure*}[!h]
\centering
\includegraphics[scale=0.6]{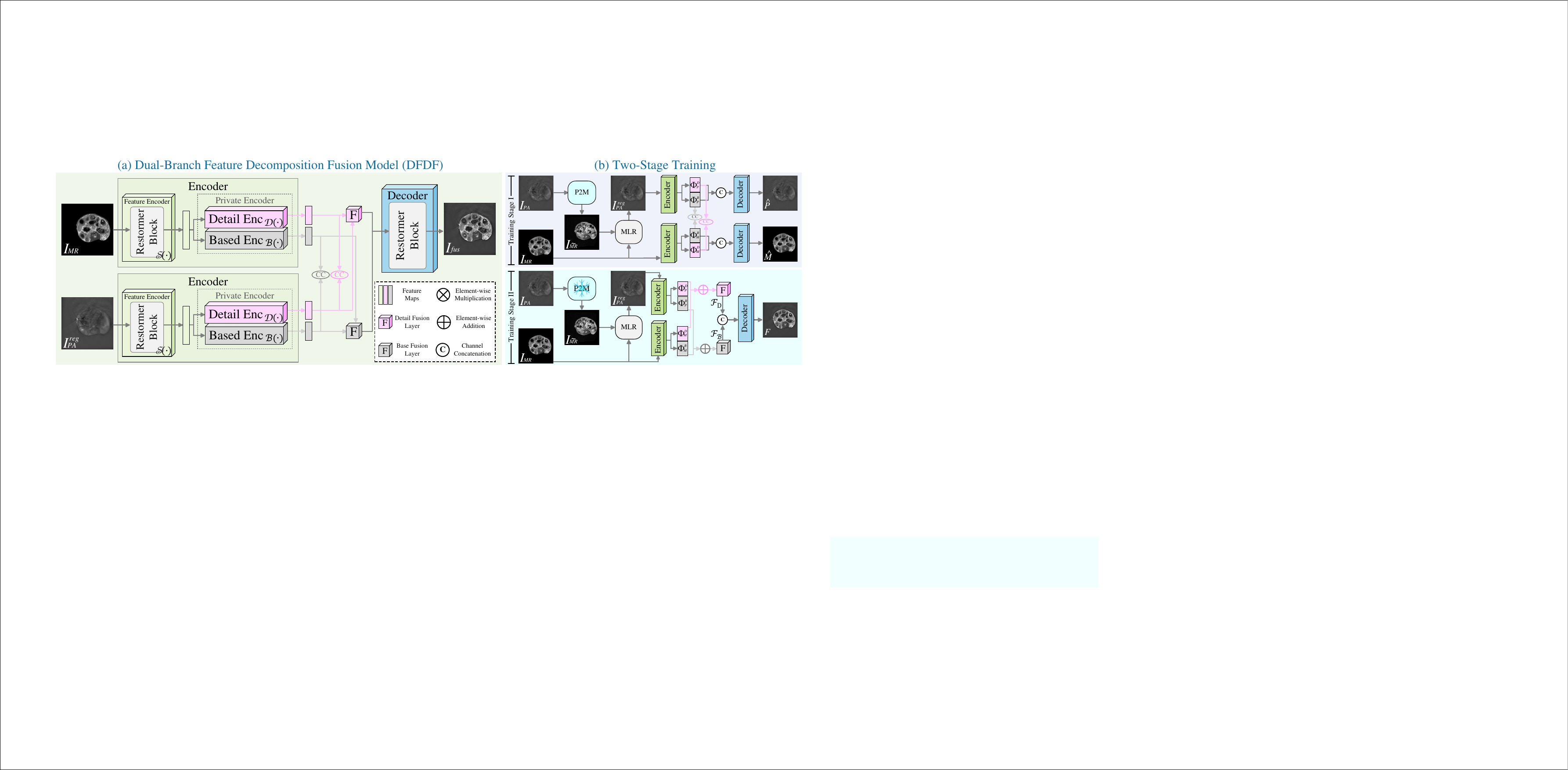}
\caption{(a) The workflow of the proposed DFDF module, which consists of three parts: encoder, fusion layer, and decoder. (b) Two-stage training learning schem.}
\end{figure*}

\subsection{Dual-Branch Feature Decomposition Fusion}
Our work explores a rational paradigm to address the challenges in PAT-MRI feature extraction and fusion. We assume that the input features of the two modalities are correlated at low frequencies, representing shared information between modalities, while high-frequency features are uncorrelated, representing the unique features of each modality. Specifically, since PAT and MRI images come from the same imaging object, the low-frequency information of both modalities includes overall tissue morphology and approximate organ positions. In contrast, the high-frequency information of both modalities is independent, such as optical absorption information in PAT images and soft tissue structural information in MRI images. Therefore, our objective is to promote the extraction of modality-specific features and shared features by respectively increasing and decreasing the correlation between low-frequency and high-frequency features. For simplicity, we represent low-frequency features as base features and high-frequency features as detail features. The specific workflow is illustrated in Fig. 3 (a). 

\subsubsection{Encoder}
The dual-branch encoder consists of three components: the feature encoder (FE) based on Restormer blocks \cite{b32}, the base encoder (BE) based on lite transformer blocks (LT) \cite{b33}, and the detail encoder (DE) based on invertible neural network (INN) blocks \cite{b34}. The BE and DE together constitute the long-short distance encoder. The paired PAT and MRI images are denoted as $P \in \mathbb{R}^{H \times W}$ and $M \in \mathbb{R}^{H \times W}$ respectively. The FE, BE, and DE are represented by $\mathcal{S}( \cdot )$, $\mathcal{B}( \cdot )$, and $\mathcal{D}( \cdot )$ respectively.

\paragraph{Feature encoder}
The objective of FE is to extract shallow features $\left\{ {\Phi_{P}^{S},\Phi_{M}^{S}} \right\}$ from the input PAT and MRI images $\left\{ {P,M} \right\}$. 

\begin{equation} 
\Phi_{P}^{S} = \mathcal{S}(P),~\Phi_{M}^{S} = \mathcal{S}(M).
\end{equation}

\paragraph{Base encoder}
BE extracts low-frequency basic features from shared features: 

\begin{equation} 
\Phi_{P}^{B} = \mathcal{B}\left( \Phi_{P}^{S} \right),~\Phi_{M}^{B} = \mathcal{B}\left( \Phi_{M}^{S} \right).
\end{equation}

\noindent The basic features of P and M are denoted as $\Phi_{P}^{B}$ and $\Phi_{M}^{B}$ respectively.  

\paragraph{Detail encoder}
In contrast to BE, DE extracts high-frequency detailed information from shared features: 

\begin{equation} 
\Phi_{P}^{D} = \mathcal{D}\left( \Phi_{P}^{S} \right),~\Phi_{M}^{D} = \mathcal{D}\left( \Phi_{M}^{S} \right).
\end{equation}

\subsubsection{Fusion layer}
The function of the base/detail fusion layer is to merge the basic/detail features separately. We use LT and INN blocks as the basic and detail fusion layers, respectively, where: 

\begin{equation} 
\Phi^{B} = \mathcal{F}_{\mathcal{B}}\left( {\Phi_{P}^{B},\Phi_{M}^{B}} \right),~\Phi^{D} = \mathcal{F}_{D}\left( {\Phi_{P}^{D},\Phi_{M}^{D}} \right).
\end{equation}

\noindent $\mathcal{F}_{\mathcal{B}}$ and $\mathcal{F}_{\mathcal{D}}$ are the basic fusion layer and detail fusion layer, respectively. 

\subsubsection{Decoder}
In the decoder $\mathcal{D}\mathcal{C}( \cdot )$, the decomposed features are concatenated along the channel dimension as input, and the original image (during training phase I) or the fused image (during training phase II) is the output of the decoder, expressed by the following formula:

\begin{equation} 
\begin{aligned} 
&Stage~I:~\hat{I} = \mathcal{D}\mathcal{C}\left( {\Phi_{P}^{B},\Phi_{P}^{D}} \right),~\hat{V} = \mathcal{D}\mathcal{C}\left( {\Phi_{M}^{B},\Phi_{M}^{D}} \right); \\
&Stage~II:~F = \mathcal{D}\mathcal{C}\left( {\Phi^{B},\Phi^{D}} \right).
\end{aligned} 
\end{equation}

\subsection{Two-stage training}
Existing multi-task model optimization paradigms either utilize pretrained high-level models to guide the training of low-level task models or jointly train multi-task models in a single stage. However, in the field of image fusion, it is challenging to provide ideal fusion images for training models. Additionally, single-stage joint training strategies may lead to difficulties in maintaining performance balance among multi-task models. Therefore, we have designed a two-stage training learning scheme for joint adaptive training of our network, as illustrated in Fig. 3 (b).

\subsubsection{Training stage I}
Firstly, unaligned PAT and MRI images $\left\{ {I_{PA},I_{MR}} \right\}$ are input into the synthesis network P2M to generate a pseudo MRI image $I_{\overset{\sim}{MR}}$. The registration network MLR performs fine registration using the pseudo MRI and MRI images $\left\{ {I_{\overset{\sim}{MR}},I_{MR}} \right\}$ as inputs, generating multiscale affine parameters and performing spatial transformation on $I_{PA}$. Pre-aligned PAT and MRI images $\left\{ {P,M} \right\}$ are input into FE to extract shallow features $\left\{ {\Phi_{P}^{S},\Phi_{M}^{S}} \right\}$. Then, low-frequency basic features $\left\{ {\Phi_{P}^{B},\Phi_{M}^{B}} \right\}$ and high-frequency detail features 
$\left\{ {\Phi_{P}^{D},\Phi_{M}^{D}} \right\}$ of the two different modalities are extracted using BE and DE, respectively. Subsequently, the basic and detail features of PAT $\left\{ {\Phi_{P}^{B},\Phi_{P}^{D}} \right\}$ (or MRI $\left\{ {\Phi_{M}^{B},\Phi_{M}^{D}} \right\}$) are concatenated and input into the decoder to reconstruct the original PAT image $\hat{P}$ (or MRI image $\hat{M}$).

\subsubsection{Training stage II}
We freeze the parameters of the well-trained image synthesis network P2M obtained in training stage I. Paired but unaligned PAT and MRI images $\left\{ {I_{PA},I_{MR}} \right\}$ are input into the image synthesis network trained in training stage I to obtain paired pseudo MRI and MRI images $\left\{ {I_{\overset{\sim}{MR}},I_{MR}} \right\}$. The pseudo MRI and MRI images are registered in the image registration network MLR, undergoing spatial transformation and correcting local misalignments, resulting in the registered PAT image $I_{PA}^{reg} = P$. Then, the registered PAT and MRI images $\left\{ {P,M} \right\}$ are input into the trained encoder to obtain decomposed features. The decomposed basic features $\left\{ {\Phi_{P}^{B},\Phi_{M}^{B}} \right\}$ and detail features $\left\{ {\Phi_{P}^{D},\Phi_{M}^{D}} \right\}$ are then input into the trained fusion layers $\mathcal{F}_{\mathcal{B}}$ and $\mathcal{F}_{\mathcal{D}}$, respectively. Finally, the fused features $\left\{ {\Phi^{B},\Phi^{D}} \right\}$ are input into the decoder to obtain the fused image $F$. 

\subsection{Loss functions}

\subsubsection{Perceptual style transfer loss}
We introduce the perceptual style transfer (PST) loss to control the cycle consistency of P2M. The PST loss consists of two terms: the perceptual loss $\mathcal{L}_{pcp}$ and the style loss $\mathcal{L}_{sty}$. First, $\mathcal{L}_{pcp}$ is defined as:

\begin{equation} 
\begin{aligned} 
\mathcal{L}_{pcp}^{\psi} = &\left\| {\psi\left( I_{PA} \right) - \psi\left( {G_{B}\left( {G_{A}\left( I_{PA} \right)} \right)} \right)} \right\|^{2} \\
+ &\left\| {\psi\left( I_{MR} \right) - \psi\left( {G_{A}\left( {G_{B}\left( I_{MR} \right)} \right)} \right)} \right\|^{2}
\end{aligned} 
\end{equation}

\noindent where $\psi$ is the layer of the VGG-19 model \cite{b36}. $\mathcal{L}_{sty}$ is defined as:

\begin{equation} 
\begin{aligned} 
\mathcal{L}_{sty}^{\psi} = &{\omega\left\| {\mathcal{G}_{\psi}\left( I_{PA} \right) - \mathcal{G}_{\psi}\left( {G_{B}\left( {G_{A}\left( I_{PA} \right)} \right)} \right)} \right\|}^{2} \\
+ &{\omega\left\| {\mathcal{G}_{\psi}\left( I_{MR} \right) - \mathcal{G}_{\psi}\left( {G_{A}\left( {G_{B}\left( I_{MR} \right)} \right)} \right)} \right\|}^{2}
\end{aligned} 
\end{equation}

\noindent where $\mathcal{G}$ is the Gram matrix \cite{b49}. The PST loss is given by: 

\begin{equation} 
\mathcal{L}_{pst} = \lambda_{p}\mathcal{L}_{pcp} + \lambda_{s}\mathcal{L}_{sty}
\end{equation}

\noindent where $\lambda_{p}$ is set to 1, and $\lambda_{s}$ is set to 100. 

\subsubsection{Registration loss}
We employ bidirectional similarity loss to constrain the registration between MRI images and pseudo-MRI images, defined as:

\begin{equation} 
\begin{aligned} 
\mathcal{L}_{sim}^{bi} = &\left\| {\psi\left( I_{MR}^{reg} \right) - \psi\left( I_{\overset{\sim}{MR}} \right)} \right\|  \\
+ \lambda_{rev}&\left\| {\psi\left( {\phi{^\circ}I_{\overset{\sim}{MR}}} \right) - \psi\left( I_{MR} \right)} \right\|_{1}
\end{aligned} 
\end{equation}

\noindent where $\lambda_{rev} = 0.2$. The deformation field $\phi$ is used to warp the pseudo-MRI image $I_{\overset{\sim}{MR}}$, making it close to the warped input $I_{MR}$. 

To ensure smooth deformation fields, we define the smoothness loss as:

\begin{equation} 
\mathcal{L}_{smooth} = \left\| {\nabla\phi} \right\|_{1}
\end{equation}

Then, the registration loss is calculated by the following equation:

\begin{equation} 
\mathcal{L}_{reg} = \mathcal{L}_{sim} + {\lambda_{sm}\mathcal{L}}_{smooth}
\end{equation}

\noindent where $\lambda_{sm}$ is set to 10 in our work.

In training stage I, we jointly train our synthesis and registration networks by minimizing the following loss function:

\begin{equation} 
\mathcal{L}_{total} = \mathcal{L}_{pst} + \mathcal{L}_{GAN} + \mathcal{L}_{reg}
\end{equation}

\noindent where $\mathcal{L}_{GAN}$ inherits the functionality of CycleGAN \cite{b31} to discriminate between fake and real MRI images.

\subsubsection{Fusion loss}
In training stage I, the total loss $\mathcal{L}_{total}^{I}$ of the image fusion network is calculated as:

\begin{equation} 
\mathcal{L}_{total}^{I} = \mathcal{L}_{PA} + \alpha_{1}\mathcal{L}_{MR} + \alpha_{2}\mathcal{L}_{decomp}
\end{equation}

\noindent where $\mathcal{L}_{PA}$ and $\mathcal{L}_{MR}$ denote the reconstruction losses of PAT and MRI images, $\mathcal{L}_{decomp}$ represents the feature decomposition loss, and $\alpha_{1}$ and $\alpha_{2}$ are tuning parameters.

\begin{equation} 
\mathcal{L}_{PA} = \mathcal{L}_{int}^{I}\left( {P,\hat{P}} \right) + \mu\mathcal{L}_{SSIM}\left( {P,\hat{P}} \right)
\end{equation}

\noindent The term $\mathcal{L}_{int}^{I}\left( {P,\hat{P}} \right) = \left\| {P - \hat{P}} \right\|_{2}^{2}$ and $\mathcal{L}_{SSIM}\left( {P,\hat{P}} \right) = 1 - SSIM\left( {P,\hat{P}} \right)$. Similarly, $\mathcal{L}_{MR}$ can be obtained in the same manner. 

The proposed feature decomposition loss $\mathcal{L}_{decomp}$ is defined as: 

\begin{equation} 
\mathcal{L}_{decomp} = \frac{\left( \mathcal{L}_{CC}^{D} \right)^{2}}{\mathcal{L}_{CC}^{B}} = \frac{\left( {CC\left( {\Phi_{P}^{D},\Phi_{M}^{D}} \right)} \right)^{2}}{\left( {CC\left( {\Phi_{P}^{B},\Phi_{M}^{B}} \right)} \right) + \epsilon}
\end{equation}

\noindent where $CC\left( {\cdot , \cdot} \right)$ denotes the correlation coefficient operator, and $\epsilon$ is set to 1.01 to ensure that this term is always positive. 

In training stage II, we freeze the image synthesis network and jointly optimize the registration and fusion networks. The total loss is given by:

\begin{equation} 
\mathcal{L}_{total}^{II} = \mathcal{L}_{reg} + \mathcal{L}_{int}^{II} + \alpha_{3}\mathcal{L}_{grad} + \alpha_{4}\mathcal{L}_{decomp}
\end{equation}

\noindent where $\mathcal{L}_{int}^{II} = \frac{1}{HW}\left\| {I_{f} - max\left( {I_{PA},I_{MR}} \right)} \right\|_{1}$ and $\mathcal{L}_{grad} = \frac{1}{HW}\left\| {\left| {\nabla I_{f}} \right| - max\left( {\left| {\nabla I}_{PA} \right|,\left| {\nabla I_{MR}} \right|} \right)} \right\|_{1}$. Here, $\nabla$ represents the Sobel gradient operator. $\alpha_{3}$ and $\alpha_{4}$ are tuning parameters. 

\begin{figure*}[h]
\centering
\includegraphics[scale=.23]{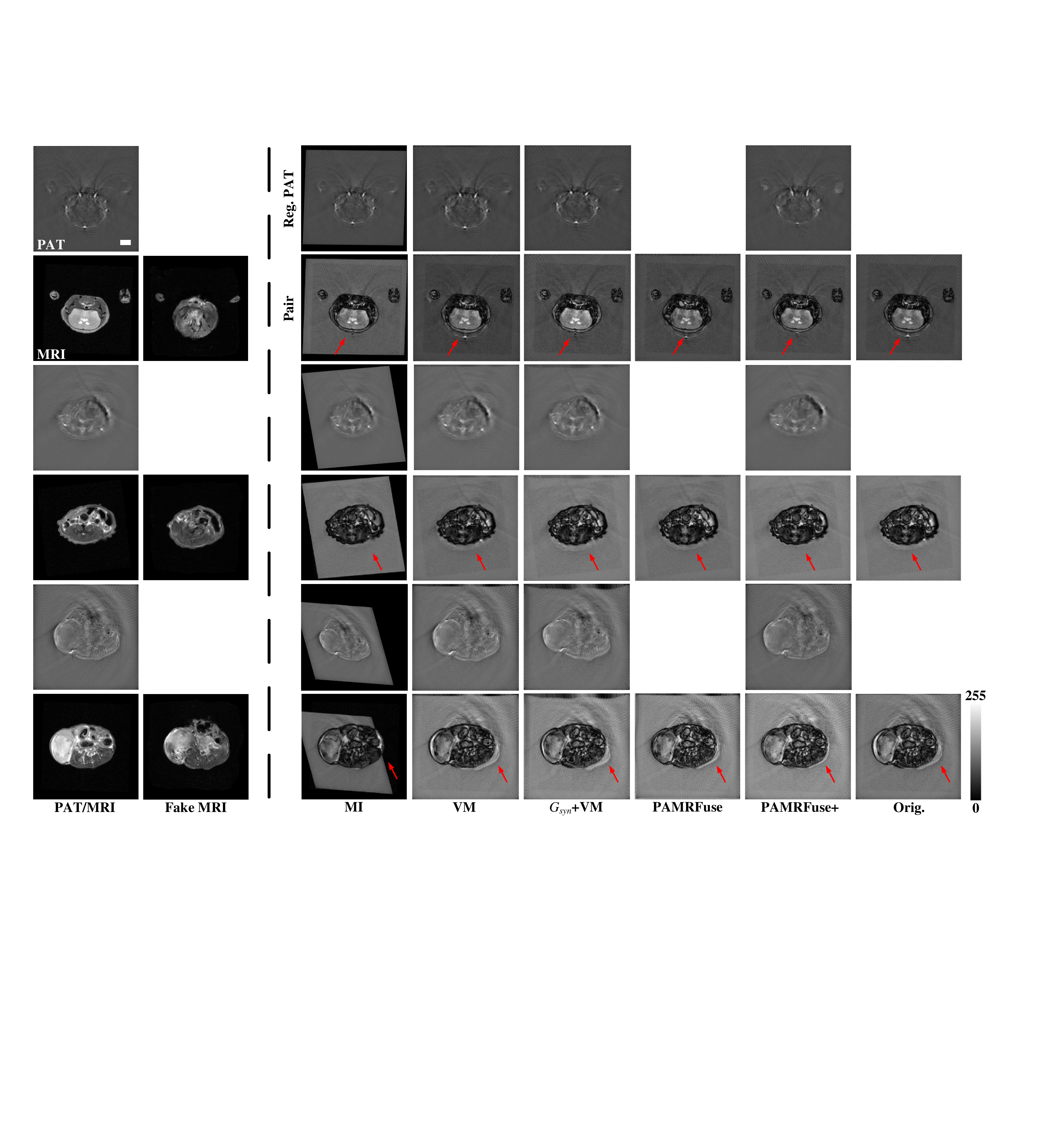}
\caption{Synthesis and registration results of PAI-MRI image pairs. Under the registration results, the distorted MRI images are shown in pairs with the PAI images to demonstrate their misalignment. Reg. = Registration. Orig.: Unaligned MRI overlayed on PAT. Scale bar, 3mm.}
\end{figure*}

\section{Experimental setup}
\subsection{Data acquisition}
For PAT imaging, a commercial small animal multispectral optoacoustic tomography system (MSOT inVision128, iThera Medical, Germany) is used. The system features tunable (660-960 nm) lasers with a pulse width of about 5 ns and a repetition rate of 10 Hz. Ultrasonic waves generated by stimulated samples are coupled through water and transmitted to a ring array transducer consisting of 128 elements. All MRI scans are performed on a 7T small animal MRI system (Pharmascan, Bruker, Germany). Furthermore, during data acquisition, we utilize a dual-modal animal imaging bed previously reported by our group \cite{b37}.

\subsection{Animal experiment}
All animal experiments have received approval from the local animal ethics committee of Southern Medical University and are conducted according to current guidelines. \textit{In vivo} animal imaging experiments involve six healthy female nude mice (12-15 g each, Southern Medical University, Guangzhou, China) and four 4T1 breast cancer nude mice (Southern Medical University Cancer Institute, Guangzhou, China). 

\subsection{Implementation details}
The algorithm \footnote{https://github.com/zhongniuniu/PAMRFuse-plus} is implemented in Python, utilizing the PyTorch framework. It runs on an Ubuntu system with an Intel Xeon E5-2667 CPU (3.2 GHz) and NVIDIA TITAN X Pascal GPU. During preprocessing, images are resized to 300×300 pixels. Data augmentation techniques including rotation and flipping result in a final dataset of 2384 pairs of PAT-MRI training data. Training is conducted for 120 epochs, with 40 epochs in the first stage and 80 epochs in the second stage. The Adam optimizer is used for model optimization with an initial learning rate of $10^{- 3}$, reduced by a factor of 0.5 every 20 epochs. For the loss functions in Eq. (16) and (19), $\alpha_{1}$ to $\alpha_{4}$ are set to 1, 2, 10, and 2 respectively.

\section{Experimental results}
\subsection{PAT-MRI image synthesis and registration}
We compare PAMRFuse+ with classical multi-modal registration methods, including mutual information (MI) \cite{b38}, Voxelmorph (VM) \cite{b39} and PAMRFuse \cite{b10}.

\subsubsection{Qualitative analysis}
The synthesis and registration results of PAT and MRI are shown in Fig. 4. MI can successfully handle some PAT-MRI misalignment scenarios and correct partial deformations, as shown in the first group of Fig. 4, but severe geometric distortions are prevalent in most image pairs, as readily observed in the last group. While PAT after the original VM registration does not exhibit severe geometric distortions, it fails to align PAT-MRI images and may even exacerbate deformations. Our PAMRFuse+ method, before registration, utilizes the image synthesis network P2M to synthesize fake MRI images from PAT. Example synthesis results as shown in the second column of Fig. 4, demonstrate that the structure depicted in the synthesized MRI images is consistent with real MRI images. More importantly, pseudo-MRI images eliminate the influence of severe scattering artifacts in the PAT image background on the registration process. Both PAMRFuse and PAMRFuse+ can correct misalignments, so their results exhibit minimal deviations. However, in the boundary blur regions of the PAT images, PAMRFuse manifested in lower registration accuracy. The effectiveness of our registration method is evident from the improvement in the regions indicated by the red arrows in Fig. 4. 

\subsubsection{Quantitative results}
We used three commonly used metrics to evaluate the registration results, including mutual information (MI) \cite{b40}, normalized mutual information (NMI), and normalized cross-correlation (NCC) \cite{b41}. As shown in Table \uppercase\expandafter{\romannumeral1}, the improvement generated by the MI algorithm in performing cross-modal alignment of PAT-MRI images can be negligible, with the CC metric even performing worse than the unaligned input. Meanwhile, the results of direct registration using the VM algorithm show that all metrics are worse compared to the unaligned input. In contrast, our PAMRFuse+ achieves improvements in all three metrics compared to the unaligned input and achieves the best performance. Compared to PAMRFuse, PAMRFuse+ is also improved in all metrics. These results demonstrate that the proposed PAMRFuse+ is more effective than existing multimodal image alignment methods.

\begin{table}[!h]
\caption{\label{tab1}Quantitative comparison of registration accuracy. \textbf{Bold} indicates the best value.}
\centering
\setlength{\tabcolsep}{3mm}{
\scalebox{0.7}{
\begin{tabular}{lcccccc}
\bottomrule[0.4mm]
\textbf{Methods}         & \textbf{MI↑}          & \textbf{NMI↑}         & \textbf{CC↑}          \\ \hline
Misaligned Input        & 0.2807$\pm$0.0494       & 0.1119$\pm$0.0146       & 0.2096$\pm$0.1239       \\ \hdashline
MI                       & 0.3016$\pm$0.0525       & 0.1194$\pm$0.0151       & 0.1947$\pm$0.0677       \\ \hdashline
VM               & 0.2696$\pm$0.0365       & 0.1071$\pm$0.0101       & 0.1904$\pm$0.0957       \\ 
P2M+VM     & $0.2923_{\textcolor{red}{\uparrow0.0227}}$$\pm$0.0464 & $0.1154_{\textcolor{red}{\uparrow0.0083}}$$\pm$0.0124 & $0.2176_{\textcolor{red}{\uparrow0.0272}}$$\pm$0.1216 \\ \hdashline
PAMRFuse               & 0.3029$\pm$0.0303       & 0.1205$\pm$0.0103       & 0.2802$\pm$0.0716       \\ 
\textbf{PAMRFuse+}      & \textbf{0.3082$\pm$0.0497}       & \textbf{0.1225$\pm$0.0145}       & \textbf{0.2900$\pm$0.0957}       \\ \bottomrule[0.4mm]
\end{tabular}}}
\end{table}

\begin{figure*}[!h]
\centering
\includegraphics[scale=.24]{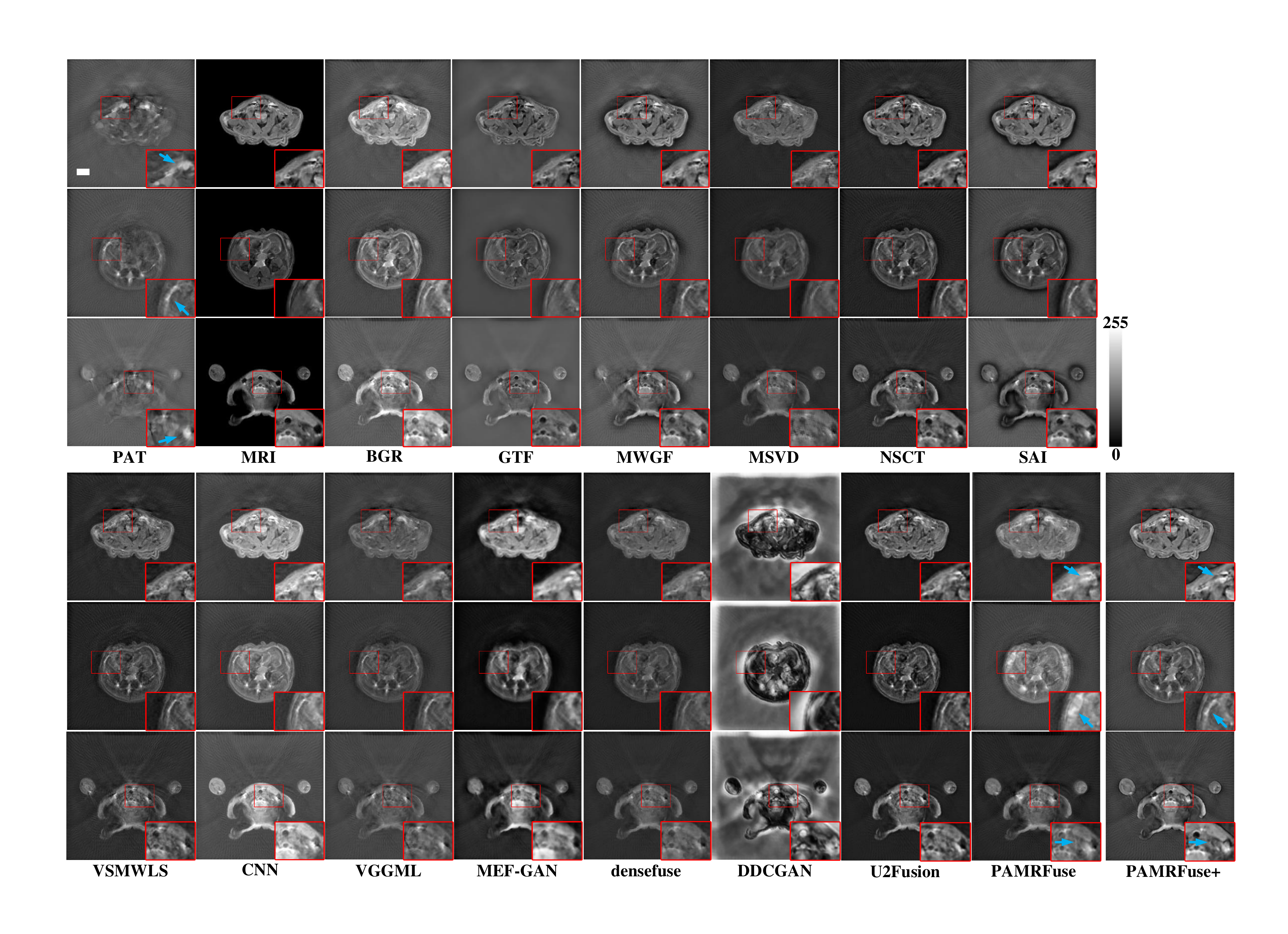}
\caption{Fusion results of unaligned PAI-MRI image pairs. Scale bar, 3mm.}
\end{figure*}

\subsection{PAT-MRI image fusion}
We compared PAMRFuse+ with state-of-the-art (SOTA) fusion methods. BGR \cite{b42}, GTF \cite{b16}, MWGF \cite{b43}, MSVD \cite{b44}, NSCT \cite{b45}, SAI \cite{b46}, and VSMWLS \cite{b17} are traditional image fusion methods. CNN \cite{b47}, VGGML \cite{b48}, densefuse \cite{b18}, U2Fusion \cite{b50}, and PAMRFuse \cite{b10} are end-to-end image fusion methods, while MEF-GAN \cite{b49} and DDCGAN \cite{b22} are GAN-based image fusion methods.

\begin{table*}[!h]
\caption{\label{tab2}Quantitative comparison of fusion performance. \textbf{Bold} indicates the best value.}
\centering
\setlength{\tabcolsep}{3mm}{
\scalebox{0.75}{
\begin{tabular}{lccccccccc}
\bottomrule[0.4mm]
             & \textbf{MI↑}     & \textbf{VIF↑}    & \textbf{Qabf↑}    & \textbf{SF↑}     & \textbf{SSIM↑}    & \textbf{$\mathbf{FMI_{pixel}}$↑}    & \textbf{$\mathbf{FMI_{dct}}$↑}    & \textbf{$\mathbf{FMI_w}$↑}  & \textbf{CC↑}    \\ \hline
\multicolumn{5}{l}{\textbf{Aligned Method: P2M + Voxelmorph}} \\ \hline
BGR          & 2.8405$\pm$0.3867  & 0.7860$\pm$0.1305  & 0.5202$\pm$0.0293  & 0.0402$\pm$0.0063  & 0.9284$\pm$0.0156  & 0.8917$\pm$0.0136  & 0.2080$\pm$0.0178  & 0.2031$\pm$0.0163 & 0.7119$\pm$0.0516\\
GTF          & 1.7997$\pm$0.3419  & 0.5441$\pm$0.1091  & 0.3401$\pm$0.0430  & 0.0301$\pm$0.0059  & 0.8478$\pm$0.0370  & 0.8904$\pm$0.0134  & 0.2317$\pm$0.0164  & 0.1493$\pm$0.0239 & 0.5836$\pm$0.0903\\
MWGF      & 2.0958$\pm$0.2824  & 0.7941$\pm$0.1521  & 0.5440$\pm$0.0346  & 0.0390$\pm$0.0065  & 0.9048$\pm$0.0197  & 0.9025$\pm$0.0117  & 0.2127$\pm$0.0201  & 0.1975$\pm$0.0166 & 0.3824$\pm$0.1681\\
MSVD      & 2.4940$\pm$0.3785  & 0.6187$\pm$0.1185  & 0.2753$\pm$0.0566  & 0.0283$\pm$0.0060  & 0.8884$\pm$0.0226  & 0.8822$\pm$0.0148  & 0.1896$\pm$0.0148  & 0.1668$\pm$0.0201 & 0.7390$\pm$0.0478\\
NSCT      & 2.2099$\pm$0.2545  & 0.6541$\pm$0.1277  & 0.5004$\pm$0.0322  & 0.0395$\pm$0.0062  & 0.9407$\pm$0.0072  & 0.8987$\pm$0.0120  & 0.2156$\pm$0.0159  & 0.1983$\pm$0.0158 & 0.7169$\pm$0.0530\\
SAI       & 2.2028$\pm$0.3645  & 0.7558$\pm$0.1539  & 0.5456$\pm$0.0369  & 0.0384$\pm$0.0065  & 0.8946$\pm$0.0273  & 0.8996$\pm$0.0117  & 0.2065$\pm$0.0183  & 0.1975$\pm$0.0174 & 0.3772$\pm$0.1618\\
VSMWLS    & 2.3324$\pm$0.2688  & 0.7026$\pm$0.1049  & 0.4307$\pm$0.0322  & 0.0354$\pm$0.0061  & 0.9256$\pm$0.0131  & 0.8828$\pm$0.0137  & 0.2051$\pm$0.0139  & 0.1980$\pm$0.0156 & 0.7330$\pm$0.0450\\
CNN       & 2.1382$\pm$0.3233  & 0.8323$\pm$0.1961  & 0.5446$\pm$0.0322  & 0.0393$\pm$0.0060  & 0.9421$\pm$0.0100  & 0.8980$\pm$0.0123  & 0.2073$\pm$0.0162  & 0.1996$\pm$0.0163 & 0.7007$\pm$0.0575\\
VGGML     & 2.5039$\pm$0.3963 & 0.6983$\pm$0.1027 & 0.3635$\pm$0.0556 & 0.0250$\pm$0.0034 & 0.8946$\pm$0.0157 & 0.8910$\pm$0.0143 & 0.1861$\pm$0.0077 & 0.1838$\pm$0.0171 & 0.7623$\pm$0.0346 \\
MEF-GAN   & 2.7033$\pm$0.4111 & 0.7265$\pm$0.1613 & 0.3325$\pm$0.0745 & 0.0289$\pm$0.0051 & 0.9078$\pm$0.0185 & 0.8829$\pm$0.0144 & 0.1908$\pm$0.0185 & 0.2433$\pm$0.0270 & 0.7147$\pm$0.0540 \\
densefuse & 3.1141$\pm$0.3915 & 0.6148$\pm$0.1173 & 0.3345$\pm$0.0176 & 0.0237$\pm$0.0035 & 0.9273$\pm$0.0121 & 0.8875$\pm$0.0138 & 0.2881$\pm$0.0120 & 0.2922$\pm$0.0136 & 0.7449$\pm$0.0470 \\
DDCGAN    & 1.6878$\pm$0.3415 & 0.0417$\pm$0.0113 & 0.2661$\pm$0.0350 & 0.0418$\pm$0.0067 & 0.3866$\pm$0.0718 & 0.8629$\pm$0.0137 & 0.2248$\pm$0.0109 & 0.1671$\pm$0.0400 & 0.4621$\pm$0.1390 \\
U2Fusion  & 3.0122$\pm$0.4112 & 0.6378$\pm$0.1589 & 0.4319$\pm$0.0468 & 0.0375$\pm$0.0048 & 0.9549$\pm$0.0100 & 0.8875$\pm$0.0143 & 0.2981$\pm$0.0178 & 0.2925$\pm$0.0199 & 0.7273$\pm$0.3880 \\
PAMRFuse  & 2.2662$\pm$0.4533 & 0.7391$\pm$0.1879 & 0.3115$\pm$0.0726 & 0.0315$\pm$0.0058 & 0.8152$\pm$0.0549 & 0.8977$\pm$0.0119 & 0.2404$\pm$0.0122 & 0.2305$\pm$0.0527 & \textbf{0.7907$\pm$0.0372} \\
\textbf{PAMRFuse+}  & \textbf{4.2421$\pm$0.3474}  & \textbf{0.8421$\pm$0.1462}  & \textbf{0.7099$\pm$0.0340}  & \textbf{0.0434$\pm$0.0063}  & \textbf{0.9629$\pm$0.0096}  & \textbf{0.9077$\pm$0.0107} & \textbf{0.3040$\pm$0.0214}  & \textbf{0.4498$\pm$0.0217}  & 0.6877$\pm$0.0627 \\
\bottomrule[0.4mm]
\end{tabular}}}
\end{table*}

\begin{figure*}[h]
\centering
\includegraphics[scale=.27]{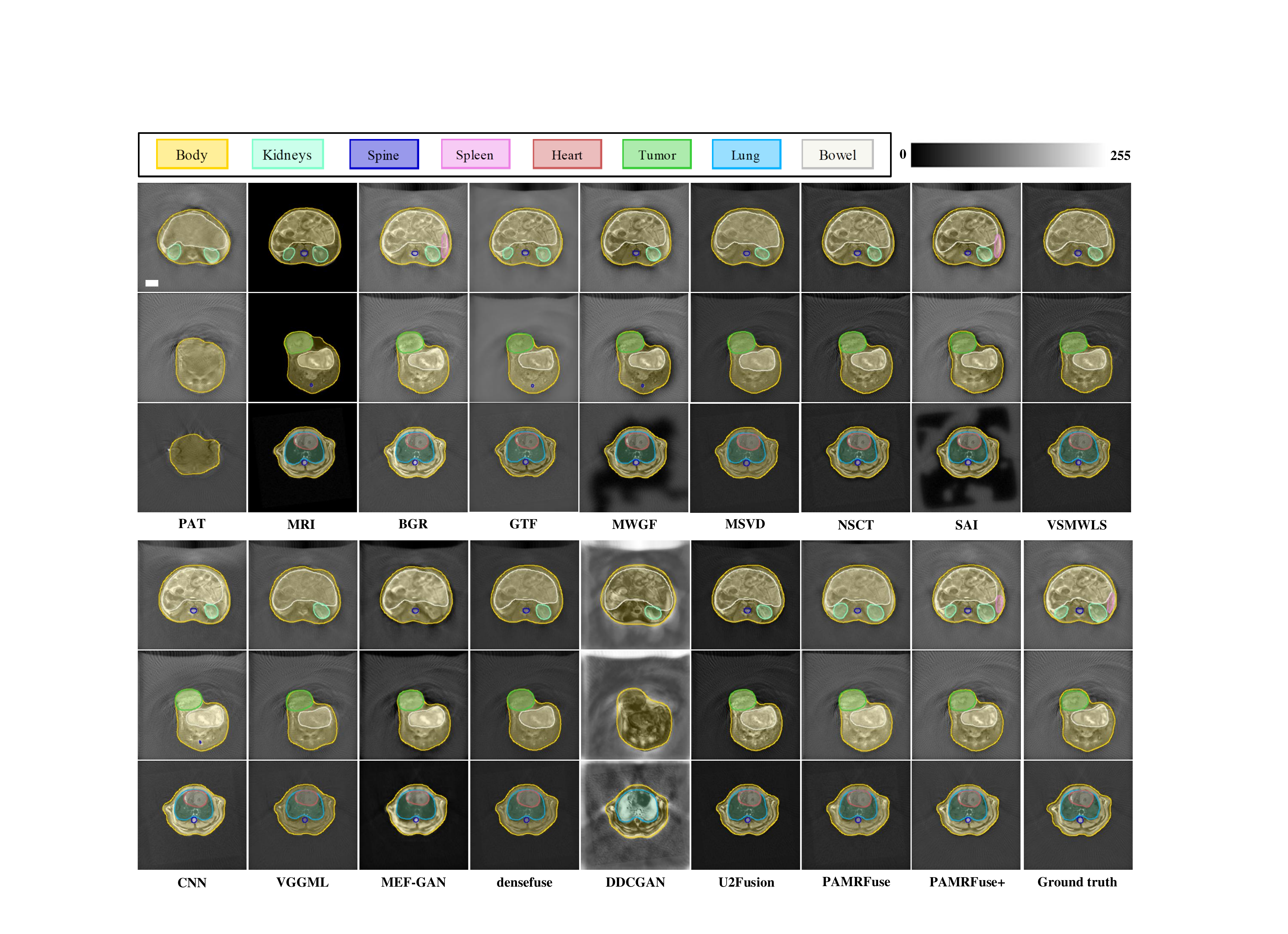}
\caption{Results of multi-organ segmentation. scale bar, 3mm.}
\end{figure*}

\subsubsection{Qualitative analysis}
Except for PAMRFuse, most of the SOTA fusion methods cannot handle unaligned data. Therefore, for a fair comparison, the registration method P2M+VM is used as their pre-registration operation. Qualitative results are shown in Fig. 5. In the first and second sets of PAT-MRI image pairs in Fig. 5, the source images are almost aligned, so we focus on comparing fusion performance. In the competitors, there are noticeable color distortions in BGR, CNN, MEF-GAN, and DDCGAN. MSVD and U2Fusion methods can extract sufficient spatial details from the source images, but the fused images produce adverse artifacts not present in the source images. NSCT, VSMWLS, and MEF-GAN methods effectively prevent visual artifacts but decrease brightness and contrast in certain areas of the fused images. The main drawbacks of MWGF and VGGML methods lie in small details from the source images observed to be blurred in the fused images. GTF and densefuse overly preserve the texture of MRI images, weakening the functional information in PAT images. In PAMRFuse, the blue-arrowed areas in Fig. 5 appear blurred, whereas in PAMRFuse+ they are clearer and more detailed. In the third set of PAT-MRI images, significant non-rigid misalignments are present in the source images. By observing locally magnified regions, we can see that misalignments still exist in the competitors' results, resulting in overlapping shadows and blurred textures. In contrast, PAMRFuse+ demonstrates good alignment and fusion capabilities.

\subsubsection{Quantitative results}
Table \uppercase\expandafter{\romannumeral2} presents a quantitative comparison of the fusion results of PAT-MRI. Compared to the other 13 methods, our approach significantly outperforms existing fusion methods, ranking first in eight metrics including MI, VIF \cite{b51}, Qabf \cite{b52}, SF \cite{b53}, SSIM, $\rm FMI_{pixel}$, $\rm FMI_{dct}$, and $\rm FMI_w$ \cite{b54}. PAMRFuse achieves the highest CC value, indicating a better retention of original information. However, PAMRFuse+ offers richer texture details, higher image quality, and improved visual fidelity.

\subsection{Task-driven evaluation}

\begin{table*}[h]
\caption{\label{tab4}Quantitative evaluation of the multi-organ instance segmentation accuracy. \textbf{Bold} and underline indicate the best and second-best values, respectively.}
\centering
 \setlength{\tabcolsep}{3mm}{
 \scalebox{0.85}{
\begin{tabular}{lccccccccc}
\bottomrule[0.4mm]
\textbf{Method} &\textbf{Body}  & \textbf{Brain} & \textbf{Lung} & \textbf{Heart} & \textbf{Intestines} & \textbf{Kidneys} & \textbf{Spleen} & \textbf{Spine} & \textbf{Tumor}
      \\ \bottomrule[0.4mm]
PAT       & 66.96$\pm$37.00 & /     & /     & /     & 30.66$\pm$38.46 & 13.75$\pm$29.74 & /     & /     & /     \\
MRI       & 95.85$\pm$3.55 & \underline{95.25$\pm$1.27} & 95.97$\pm$0.59 & 93.11$\pm$4.82 & \underline{93.91$\pm$3.75} & 79.85$\pm$26.27 & 36.53$\pm$43.35 & \textbf{82.96$\pm$18.75} & 54.57$\pm$47.20 \\
BGR       & 95.28$\pm$3.88 & 93.95$\pm$1.91 & 96.48$\pm$0.31 & 84.14$\pm$16.4 & 92.23$\pm$2.94 & 72.75$\pm$28.52 & 34.30$\pm$40.91 & 70.51$\pm$27.75 & 60.09$\pm$42.19 \\
GTF       & 95.11$\pm$3.02 & 88.62$\pm$4.85 & 96.42$\pm$0.43 & 91.86$\pm$6.27 & 87.97$\pm$18.67 & 77.23$\pm$25.84 & \underline{46.06$\pm$44.00} & 76.83$\pm$24.08 & 52.68$\pm$45.83 \\
MWGF      & 88.74$\pm$24.11 & 95.14$\pm$1.61 & 96.19$\pm$0.71 & \underline{93.56$\pm$3.62} & 89.54$\pm$19.04 & 77.12$\pm$26.06 & 35.16$\pm$41.70 & 77.39$\pm$24.25 & 71.64$\pm$28.42 \\
MSVD      & 92.47$\pm$18.55 & 90.72$\pm$3.30 & 95.85$\pm$1.07 & 90.83$\pm$6.34 & 87.23$\pm$18.83 & 61.57$\pm$36.31 & /     & 74.17$\pm$28.52 & 68.80$\pm$37.26 \\
NSCT      & 95.98$\pm$4.25 & 93.75$\pm$2.11 & 96.21$\pm$0.55 & 90.28$\pm$10.32 & 92.55$\pm$3.38 & 72.84$\pm$28.48 & 18.33$\pm$36.38 & 73.90$\pm$28.61 & 69.55$\pm$37.40 \\
SAI       & 94.02$\pm$13.62 & 94.51$\pm$1.70 & 96.36$\pm$0.63 & 89.59$\pm$6.39 & 92.67$\pm$3.47 & 76.48$\pm$21.53 & 25.42$\pm$38.47 & 73.25$\pm$28.43 & 69.38$\pm$37.27 \\
VSMWLS    & 90.83$\pm$18.7 & 91.97$\pm$2.91 & 96.27$\pm$0.63 & 87.09$\pm$10.99 & 89.70$\pm$3.50 & 60.10$\pm$35.30 & /     & 71.48$\pm$32.26 & 59.86$\pm$42.02 \\
CNN       & 96.25$\pm$4.20 & 94.56$\pm$1.54 & 95.97$\pm$1.06 & 91.11$\pm$6.53 & 88.92$\pm$18.78 & 72.90$\pm$28.53 & 26.86$\pm$40.38 & 75.31$\pm$24.00 & 87.44$\pm$6.68 \\
VGGML     & 91.99$\pm$14.14 & 81.00$\pm$28.68 & 95.93$\pm$1.04 & 69.98$\pm$40.47 & 81.93$\pm$25.12 & 38.14$\pm$40.5 & 9.10$\pm$27.29  & 48.09$\pm$40.59 & 42.07$\pm$45.18 \\
MEF-GAN   & 93.98$\pm$6.06 & 91.40$\pm$2.88 & 95.44$\pm$2.12 & 82.22$\pm$5.58 & 91.57$\pm$4.86 & 43.03$\pm$42.65 & 9.78$\pm$9.33  & 68.54$\pm$31.48 & \underline{82.06$\pm$29.06} \\
densefuse & 95.86$\pm$4.15 & 93.34$\pm$1.70 & \underline{96.72$\pm$0.77} & 91.14$\pm$7.80 & 88.51$\pm$18.67 & 73.26$\pm$28.44 & 18.01$\pm$35.74 & 77.19$\pm$17.74 & 77.96$\pm$28.35 \\
DDCGAN    & 86.80$\pm$25.04 & 48.43$\pm$42.53 & 52.43$\pm$47.96 & /     & 73.16$\pm$33.00 & 4.15$\pm$16.61  & /     & 7.02$\pm$24.31  & 7.40$\pm$23.42 \\
U2Fusion  & 95.46$\pm$4.22 & 91.41$\pm$3.54 & \textbf{96.89$\pm$0.51} & 88.95$\pm$6.72 & 92.35$\pm$3.17 & 72.50$\pm$28.05 & 8.34$\pm$25.03  & 77.56$\pm$18.39 & 61.11$\pm$43.02 \\
PAMRFuse  & \underline{96.42$\pm$4.09} & 94.03$\pm$1.29 & 95.22$\pm$1.61 & 87.87$\pm$10.14 & 85.61$\pm$25.86 & \underline{82.23$\pm$16.73}  & 25.88$\pm$38.86     &71.20$\pm$28.17  & 80.89$\pm$28.77 \\
\textbf{PAMRFuse+}   & \textbf{98.04$\pm$0.72}     & \textbf{95.54$\pm$1.15}       & 96.12$\pm$0.21       & \textbf{93.60$\pm$1.23}       & \textbf{94.49$\pm$1.51}       & \textbf{88.86$\pm$5.39}       & \textbf{84.08$\pm$3.35}       & \underline{80.77$\pm$7.99}   & \textbf{88.25$\pm$6.76}   \\
\bottomrule[0.4mm]
\end{tabular}}}
\end{table*}

We employ a SOTA PAT instance segmentation model, the structure fusion enhanced graph convolutional multi-organ instance segmentation pipeline (SFE-GCN), to evaluate multi-organ instance segmentation performance on fused images. 14 instance segmentation models are trained separately on the PAT, MRI, and 14 fusion image training datasets. Visual examples in Fig. 6 illustrate the advantages of the fusion algorithm in facilitating multi-organ instance segmentation. In the first image pair, BGR, SAI, and PAMRFuse+ successfully segment the spleen. However, inappropriate fusion may weaken significant targets due to interference from negative and irrelevant information. In the first image pair, the left kidney could have been segmented in PAT and MRI source images but gets unsegmented after enhancement by most contrast fusion methods. Additionally, erroneous or insufficient fusion may weaken complementary information from source images, leading to incorrect organ segmentation. For example, in the second image pair, GTF, MWGF, and CNN erroneously segment a non-existent spine. In contrast, SFE-GCN successfully segments all target organs from our fused images, proving that our fusion images provide more semantic information for downstream tasks.

We further employ quantitative metrics to assess multi-organ instance segmentation tasks. Using the DSC metric to measure segmentation performance, results are presented in Table \uppercase\expandafter{\romannumeral3}. MRI images achieve the highest DSC values for the spine, indicating that MRI images provide sufficient semantic information about the spine for the segmentation network. However, the segmentation results for the spleen and tumors in MRI images are disappointing. Fortunately, PAT images can provide rich semantic information about the spleen and tumors for the segmentation network. In Table \uppercase\expandafter{\romannumeral3}, our fusion results show the highest DSC values for most organs. These findings demonstrate that PAMRFuse+ effectively integrates the photoacoustic signals from PAT images and the soft tissue information from MRI images, thus achieving the highest segmentation accuracy in multi-organ instance segmentation tasks.

\subsection{Ablation study}

\begin{table*}[tp]
\caption{\label{tab5}Quantitative results of ablation studies.}
\centering
\setlength{\tabcolsep}{3mm}{
\scalebox{0.85}{
\begin{tabular}{lccccccccccccc}
\bottomrule[0.4mm]
\multirow{3}{*}{\textbf{(P2M, MLR)}} & \multicolumn{12}{c}{\textbf{Metrics}} \\ 
                                     & \multicolumn{4}{c}{\textbf{Registration metrics}} & \multicolumn{8}{c}{\textbf{Fusion metrics}} \\ 
                                     & \textbf{MI} & \textbf{NMI} & \textbf{CC} & \multicolumn{2}{c}{\textbf{MI}} & \textbf{VIF} & \textbf{Qabf} & \textbf{SF} & \textbf{SSIM} & \textbf{$\mathbf{FMI_{pixel}}$} & \textbf{$\mathbf{FMI_{dct}}$} & \textbf{$\mathbf{FMI_w}$} \\ 
\hline
(\XSolidBrush, \XSolidBrush)         & \textbackslash{} & \textbackslash{} & \textbackslash{} & \multicolumn{2}{c}{1.5281} & 0.4590 & 0.3958 & 0.0438 & 0.7240 & 0.8927 & 0.2071 & 0.1500 \\ 
(\XSolidBrush, \Checkmark)             & 0.3019 & 0.1199 & 0.2826 & \multicolumn{2}{c}{1.5532} & 0.4672 & 0.3956 & 0.0428 & 0.7309 & 0.8941 & 0.2134 & 0.1515 \\ 
(\Checkmark, \Checkmark)             & \textbf{0.3082} & \textbf{0.1225} & \textbf{0.2900} & \multicolumn{2}{c}{\textbf{4.2421}} & \textbf{0.8421} & \textbf{0.7099} & \textbf{0.0434} & \textbf{0.9629} & \textbf{0.9077} & \textbf{0.3040} & \textbf{0.4498} \\ 
\bottomrule[0.4mm]
\multicolumn{13}{l}{\textbf{(feedback of fusion, two-stage training)}} \\ 
\hline
(\XSolidBrush, \XSolidBrush)         & 0.3026 & 0.1203 & 0.2796 & \multicolumn{2}{c}{1.4794} & 0.4075 & 0.3189 & 0.0377 & 0.7116 & 0.8829 & 0.1807 & 0.1438 \\ 
(\Checkmark, \XSolidBrush)             & 0.3033 & 0.1205 & 0.2813 & \multicolumn{2}{c}{1.6235} & 0.4652 & 0.3225 & 0.0372 & 0.7523 & 0.8897 & 0.2349 & 0.1638 \\ 
(\XSolidBrush, \Checkmark)             & 0.3026 & 0.1203 & 0.2796 & \multicolumn{2}{c}{1.5504} & 0.4653 & 0.3953 & 0.0427 & 0.7303 & 0.8940 & 0.2133 & 0.1513 \\ 
(\Checkmark, \Checkmark)             & \textbf{0.3082} & \textbf{0.1225} & \textbf{0.2900} & \multicolumn{2}{c}{\textbf{4.2421}} & \textbf{0.8421} & \textbf{0.7099} & \textbf{0.0434} & \textbf{0.9629} & \textbf{0.9077} & \textbf{0.3040} & \textbf{0.4498} \\ 
\bottomrule[0.4mm]
\end{tabular}}}
\end{table*}

\subsubsection{Validity of synthetic and alignment networks} 
Two key components of our fusion framework are the synthesis network P2M and the registration network MLR, which effectively collaborate to mitigate the misalignment between PAT and MRI, reducing artifacts during the fusion process. To evaluate their effectiveness, we conducted ablation experiments on P2M and MLR. In Section \uppercase\expandafter{\romannumeral4}. A, we inserted the image synthesis network P2M into VM as an enhanced version. As shown in Table \uppercase\expandafter{\romannumeral1}, the quantitative results obtained with P2M in conjunction with VM showed improvements over the original version across all metrics. 

In this subsection, we comprehensively validate the effectiveness of P2M and MLR from both the perspective of image registration and fusion. Firstly, assuming the absence of P2M, we investigate the impact of the registration network MLR on the unaligned PAT-MRI fusion task. Quantitative results in Table \uppercase\expandafter{\romannumeral4} demonstrate improvements across most fusion metrics. To further enhance the accuracy of PAT-MRI image registration, we introduce the P2M network, simplifying the multi-modal PAT-MRI registration into a single-modal MRI registration. As reported in Table \uppercase\expandafter{\romannumeral4}, the combined use of P2M and MLR significantly enhances the performance of both registration and fusion in multimodal registration.

\subsubsection{Training strategy comparison} 
 In this subsection, we conduct comparative experiments to validate the effectiveness of the registration-fusion joint training strategy and the two-stage training strategy for the fusion network. If we abandon the joint optimization of the registration and fusion networks and discard the two-stage training, it can be observed that most registration and fusion metrics in Table \uppercase\expandafter{\romannumeral4} yield the least desirable performance. When we introduce the registration-fusion joint training strategy alone leads to improvements in most registration and fusion metrics. Moreover, the degree of improvement in fusion metrics in Table \uppercase\expandafter{\romannumeral4} is greater than when solely introducing the registration-fusion joint optimization strategy, indicating the criticality of this strategy for training the fusion network. Finally, when combining the registration-fusion joint training strategy with the two-stage training strategy for the fusion network, superior registration accuracy and fusion quality are observed in both qualitative comparisons and metric evaluations. In summary, the ablation results above confirm the superiority of our overall framework which relies on the cooperation of each component.

\section{Discussion}
The purpose of our image fusion is to synthesize a fused image that not only contains prominent targets and rich texture details but also facilitates the completion of advanced visual tasks. Current medical image fusion algorithms face some pressing challenges, including source image misalignment, pre-registration operation requirement, insufficient cross-model feature extraction, and the lack of validation on advanced visual tasks. 

The proposed PAMRFuse+ addresses the limitations of existing multi-modal image registration and fusion methods and achieves them in a mutually reinforcing framework. Specifically, the proposed style transfer network P2M addresses the misalignment issue between image pairs of different modalities and the challenges of multi-modal registration, and the multi-level fine registration network MLR performs image registration in a single-modal environment. In PAT-MRI image synthesis and registration experiments, excellent performance is observed in registering unaligned multi-modal images. 

Secondly, the dual-branch feature decomposition fusion network DFDF tackles the problem of insufficient cross-modal feature extraction in current CNN-based image fusion networks The quantitative and qualitative results of fusion experiments demonstrate that our fusion network DFDF can efficiently extract mode-specific features and shared features, and decompose them intuitively through the decomposition loss. Thirdly, to maintain the performance balance among the multi-task models, we have specially designed a two-stage training learning scheme. To validate the effectiveness of our proposed method, we conducted relevant ablation experiments, i.e., exploring the effects of synthesis and alignment networks, alignment and fusion mutual enhancement strategies, and two-step fine-training strategies on the fusion effect, respectively. 

Finally, we conducted task-driven evaluation experiments to demonstrate the impact of image fusion on advanced visual tasks. The performance comparison of various fusion algorithms in multi-organ segmentation results reveals the advantages of our framework in promoting advanced visual tasks. To our knowledge, this is the first time that advanced medical image visual tasks have been used to evaluate the effectiveness of image fusion. 

\section{Conclusion}
In this paper, we propose an unsupervised PAT and MRI image fusion model for facilitating advanced visual tasks, named PAMRFuse+. Firstly, we simplify the alignment of cross-modal image pairs into single-modal registration using a generation registration paradigm. Secondly, we introduce a multi-level fine registration network to predict the displacement vector field and perform registration reconstruction. Thirdly, a dual-branch feature decomposition fusion network is developed to effectively extract unique features and shared features from each modality. Extensive experimental results demonstrate the outstanding capability of PAMRFuse+ in fusing unaligned PAT-MRI images. Importantly, task-driven evaluation experiments reveal the inherent advantages of PAMRFuse+ in promoting high-level visual tasks.

\appendices


\begin{thebibliography}{00}
\bibitem{b1}L. V. Wang and S. Hu, “Photoacoustic Tomography: \textit{In Vivo} Imaging from Organelles to Organs,” \emph{Science}, vol. 335, no. 6075, pp. 1458–1462, Mar. 2012.
\bibitem{b2}K. Huda, D. J. Lawrence, W. Thompson, S. H. Lindsey, and C. L. Bayer, “\textit{In vivo} noninvasive systemic myography of acute systemic vasoactivity in female pregnant mice,” \emph{Nat. Commun.}, vol. 14, no. 1, Oct. 2023.
\bibitem{b5}S. Zhang et al., “MRI Information-Based Correction and Restoration of Photoacoustic Tomography,” \emph{IEEE Trans. Med. Imaging}, vol. 41, no. 9, pp. 2543–2555, Sep. 2022.

\bibitem{b7}W. Ren, Xosé Luís Deán-Ben, Mark-Aurel Augath, and D. Razansky, “Development of concurrent magnetic resonance imaging and volumetric optoacoustic tomography: A phantom feasibility study,” \emph{J. Biophotonics}, vol. 14, no. 2, Feb. 2021.
\bibitem{b8}R. Ni et al., “Coregistered transcranial optoacoustic and magnetic resonance angiography of the human brain,” \emph{Opt. Lett.}, vol. 48, no. 3, pp. 648–648, Dec. 2022.

\bibitem{b9}S. Liu, W. Li, D. He, G. Wang, and Y. Huang, “SSEFusion: Salient semantic enhancement for multimodal medical image fusion with Mamba and dynamic spiking neural networks,” \emph{Inf. Fusion}, pp. 103031–103031, Feb. 2025.

\bibitem{b10}Y. Zhong et al., “Unsupervised Fusion of Misaligned PAT and MRI Images via Mutually Reinforcing Cross-Modality Image Generation and Registration,” \emph{IEEE Trans. Med. Imaging}, pp. 1–1, Jan. 2024.

\bibitem{b28}Z. Yuan and H. Jiang, “Quantitative photoacoustic tomography: Recovery of optical absorption coefficient maps of heterogeneous media,” \emph{Appl. Phys. Lett.}, vol. 88, no. 23, p. 231101, Jun. 2006.

\bibitem{b25}Z. Liang et al., “Automatic 3-D segmentation and volumetric light fluence correction for photoacoustic tomography based on optimal 3-D graph search,” \emph{Med. Image Anal.}, vol. 75, pp. 102275–102275, Jan. 2022.

\bibitem{b31}J.-Y. Zhu, T. Park, P. Isola, and A. A. Efros, “Unpaired Image-to-Image Translation Using Cycle-Consistent Adversarial Networks,” \emph{ICCV}, Oct. 2017.
\bibitem{b32}Syed Waqas Zamir et al., “Restormer: Efficient Transformer for High-Resolution Image Restoration,” Jun. 2022.
\bibitem{b33}Z. Wu, Z. Liu, J. Lin, Y. Lin, and S. Han, “Lite Transformer with Long-Short Range Attention,” \emph{ICLR}, 2020.
\bibitem{b34}L. Dinh, J. Sohl-Dickstein, and S. Bengio, “Density estimation using Real NVP,” \emph{ICLR}, Feb. 2017.
\bibitem{b36}K. Simonyan and A. Zisserman, “Very Deep Convolutional Networks for Large-Scale Image Recognition,” \emph{ICLR}, Apr. 2015.
\bibitem{b37}S. Zhang et al., “\textit{In vivo} co-registered hybrid-contrast imaging by successive photoacoustic tomography and magnetic resonance imaging,” \emph{Photoacoustics}, vol. 31, p. 100506, Jun. 2023.
\bibitem{b38}P. Viola and W. M. Wells, “Alignment by maximization of mutual information,” \emph{ICCV}, Jun. 1995.
\bibitem{b39}G. Balakrishnan, A. Zhao, M. R. Sabuncu, J. Guttag, and A. V. Dalca, “VoxelMorph: A Learning Framework for Deformable Medical Image Registration,” \emph{IEEE Trans. Med. Imaging}, vol. 38, no. 8, pp. 1788–1800, Aug. 2019.
\bibitem{b40}G. Qu, D. Zhang, and P. Yan, “Information measure for performance of image fusion,” \emph{Electron. Lett.}, vol. 38, no. 7, p. 313, 2002.
\bibitem{b41}Z. Han, X. Tang, X. Gao, and F. Z. Hu, “IMAGE FUSION AND IMAGE QUALITY ASSESSMENT OF FUSED IMAGES,” \emph{ISPRS Ann. Photogramm. Remote Sens. Spatial Inf. Sci.}, vol. XL-7/W1, pp. 33–36, Jul. 2013.
\bibitem{b42}Y. Zhang, L. Zhang, X. Bai, and L. Zhang, “Infrared and visual image fusion through infrared feature extraction and visual information preservation,” \emph{Infrared Phys. Technol.}, vol. 83, pp. 227–237, Jun. 2017.

\bibitem{b16}J. Ma, C. Chen, C. Li, and J. Huang, “Infrared and visible image fusion via gradient transfer and total variation minimization,” \emph{Inf. Fusion}, vol. 31, pp. 100–109, Sep. 2016.
\bibitem{b17}J. Ma, Z. Zhou, B. Wang, and H. Zong, “Infrared and visible image fusion based on visual saliency map and weighted least square optimization,” \emph{Infrared Phys. Technol.}, vol. 82, pp. 8–17, May 2017.

\bibitem{b18}H. Li and X.-J. Wu, “DenseFuse: A Fusion Approach to Infrared and Visible Images,” \emph{IEEE Trans. Image Process.}, vol. 28, no. 5, pp. 2614–2623, May 2019.

\bibitem{b22}J. Ma, H. Xu, J. Jiang, X. Mei, and X.-P. Zhang, “DDcGAN: A Dual-Discriminator Conditional Generative Adversarial Network for Multi-Resolution Image Fusion,” \emph{IEEE Trans. Image Process.}, vol. 29, pp. 4980–4995, 2020.

\bibitem{b43}Z. Zhou, S. Li, and B. Wang, “Multi-scale weighted gradient-based fusion for multi-focus images,” \emph{Inf. Fusion}, vol. 20, pp. 60–72, Nov. 2014.
\bibitem{b44}V. P. S. Naidu, “Image Fusion Technique using Multi-resolution Singular Value Decomposition,” \emph{Def. Sci. J.}, vol. 61, no. 5, p. 479, Sep. 2011.
\bibitem{b45}B. Yang, S. Li, and F. Sun, “Image Fusion Using Nonsubsampled Contourlet Transform,” Aug. 2007.
\bibitem{b46}Y. Liu, X. Chen, J. Cheng, H. Peng, and Z. Wang, “Infrared and visible image fusion with convolutional neural networks,” \emph{Int. J. Wavelets Multiresolution Inf. Process.}, vol. 16, no. 03, p. 1850018, May 2018.
\bibitem{b47}H. Li, X.-J. Wu, and J. Kittler, “Infrared and Visible Image Fusion using a Deep Learning Framework,” \emph{ICPR}, pp. 2705–2710, Aug. 2018.
\bibitem{b48}M. S. M. Sajjadi, B. Scholkopf, and M. Hirsch, “EnhanceNet: Single Image Super-Resolution Through Automated Texture Synthesis,” \emph{ICCV}, Oct. 2017.
\bibitem{b49}H. Xu, J. Ma, and X.-P. Zhang, “MEF-GAN: Multi-Exposure Image Fusion via Generative Adversarial Networks,” \emph{IEEE Trans. Image Process.}, vol. 29, pp. 7203–7216, 2020.
\bibitem{b50}H. Xu, J. Ma, J. Jiang, X. Guo, and H. Ling, “U2Fusion: A Unified Unsupervised Image Fusion Network,” \emph{IEEE Trans. Pattern Anal. Mach. Intell.}, vol. 44, no. 1, pp. 502–518, Jan. 2022.

\bibitem{b51}Y. Han, Y. Cai, Y. Cao, and X. Xu, “A new image fusion performance metric based on visual information fidelity,” \emph{Inf. Fusion}, vol. 14, no. 2, pp. 127–135, Apr. 2013.
\bibitem{b52}C. S. Xydeas and PetrovicV., “Objective image fusion performance measure,” \emph{Electron. Lett.}, vol. 36, no. 4, p. 308, 2000.
\bibitem{b53}A. M. Eskicioglu and P. S. Fisher, “Image quality measures and their performance,” \emph{IEEE Trans. Commun.}, vol. 43, no. 12, pp. 2959–2965, 1995.
\bibitem{b54}M. B. A. Haghighat, A. Aghagolzadeh, and H. Seyedarabi, “A non-reference image fusion metric based on mutual information of image features,” \emph{Comput. Electr. Eng.}, vol. 37, no. 5, pp. 744–756, Sep. 2011.
\end{thebibliography}
\end{document}